# Enhanced numerical models for two-component fluid flow in multiscale porous structures

J. Yang (杨金精),[1*] H. Otomo (大友 洋),[1] Hongli Fan (范宏丽),[1] Guangyuan Sun (孙光远),[1] Rafael Salazar-Tio,[1] Ganapathi Raman Balasubramanian,[1] Ashraful Islam,[1] Bernd Crouse[1] and Raoyang Zhang (张绕阳)[1]

[1]*Simulia R&D, Dassault Systèmes Americas Corp., 175 Wyman Street, Waltham, MA 02451, USA*

Multi-component fluid flow simulations in multi-scale porous structures, many parts of which are likely to be under-resolved at practical resolution, often require a high-fidelity numerical model to account for the contribution of under-resolved structures to the fluid flow. In previous studies, a numerical model was proposed for viscous and capillary forces from under-resolved regions. It successfully showed comparable absolute permeability, capillary pressure, and relative permeability when compared to an equivalent fully resolved case up to ten times higher resolution. In this study, we show further extensions of the model to handle various types of structures and to capture the detailed fluid behavior. First, we introduce the controllability of the surface tension in the pseudo-potential lattice Boltzmann model while keeping the interface thickness and the spurious current at the same level. This helps to resolve more detailed interface dynamics in under-resolved regions. Second, we develop a numerical model to capture the residual fluid component in the under-resolved structure. Since it is difficult to capture such cell-sized or less-than-cell-sized fluid components with the diffusive interface models, we try to consider them separately using local constitutive relations, such as the absolute and relative permeability and capillary pressure curves. Third, we introduce a tensorial resistivity model to handle underresolved heterogeneous structures, such as a fiber bundle. After calculating the principal axis for the resistivity using the Hessian and the gradient of the local porosity field, the tensorial resistivity is rotated in the proper direction. Through a series of benchmark test cases, including cases using practical rock geometries, these enhancements are validated and show significant accuracy improvements with respect to the transient interface dynamics, capture of local irreducible fluid components, and directionality effects of underresolved structures.

## I. INTRODUCTION

Simulating multi-component fluid flow in porous regions with complex solid structures is critical for many industrial processes, including enhanced oil recovery techniques using carbon dioxide injection[1,2], water distributions in fuel cells [3,4,5,6], in-situ copper extraction by leaching[7,8], advanced personal protective gear[9], complex carbonate rock[10] and underfill flow in flip-chip encapsulation[11]. For high-fidelity simulations, it is essential to precisely model complex solid boundaries. However, the complete capture of every detail of multi-scale porous structures is often not feasible due to computational constraints and limitations of existing models and algorithms, even though such structures are common in nature.

Previously, computational models and methodologies were developed for the multi-component multiscale flow simulation[12]. For example, in a scanned image shown in the top images of Figure 1, where black represents the pore and white represents the solid, the small-structured porous regions are underresolved and colored in gray. To account for the effects of these small structures on fluid flow, viscous and capillary forces in such regions are numerically modeled using representative physical properties such as absolute permeability $K_0(\varphi)$, capillary pressure $P_c(Sw)$, and relative permeability $K_r^\alpha(Sw)$, where $Sw$ and $\varphi$ are water saturation and porosity, respectively. An index α indicates the fluid component. The workflow can be summarized as below;

i. Through geometric analysis for the scanned images of a porous media sample, the pore type of under-resolved porous structures is identified, as shown in the middle images in Figure 1 in which a suitable representative porous structure is selected from four candidates.

ii. Find flow models for each pore type of underresolved structure. If the corresponding flow models do not exist in the library, perform a fully resolved simulation in a small domain of such pore type structure, compute the new set of physical properties of $K_0(\varphi)$, $P_c(Sw)$, and $K_r^\alpha(Sw)$, and add them to the library as shown in the lower figures in Figure 1.

iii. Define the fluid forces at each underresolved location using the specific pore type physical properties in the library, local $Sw$, and local $\varphi$ as followings[12];

$$F_{PM,\gamma}^\alpha = -P_c(Sw)\rho_r^\alpha n_\gamma H(At, |\partial_x At|) - \sum_{\bar\alpha} \frac{\rho^{\bar\alpha} \nu^{\bar\alpha}}{K_0(\varphi) K_r^{\bar\alpha}(Sw)} \rho_r^\alpha u_\gamma, \quad (1)$$

where $n_\gamma$ is the normal vector of the interface, $\nu^{\bar\alpha}$ is the viscosity of component $\bar\alpha$, $\rho_r^\alpha = \rho^\alpha / \sum_{\bar\alpha} \rho^{\bar\alpha}$ is the relative density where $\rho^\alpha$ is the density of component α, $u_\gamma$ is the fluid velocity, and $\gamma \in$

$[x, y, z]$ is the Cartesian direction. A functional $H$ of the Atwood number $At$ and its spatial derivative form a switching function such that the first term is nonzero only around the interface. This functional is desirable for the diffusive interface model to avoid the excessive force from the interface regions. The first term in Eq. (1) represents the capillary force and the second term represents the viscous force, which is similar to the resistance term in the Brinkman equation[13].

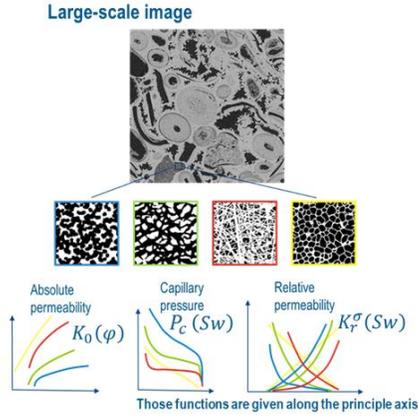

**Figure 1. Under-resolved regions in a larger scale image[14] shown with the gray-scale color. Each region is categorized with a representative porous structure. For each porous structure, the physical properties such as absolute permeability, capillary-pressure-saturation and relative permeability are computed and stored in a library for future use.**

Indeed, following this workflow and the formulation on top of the multi-phase Lattice Boltzmann model, called as the pseudo-potential model[15,16], we tested the single-phase and multi-phase flow simulation through the multi-scale porous media. By comparing the fully resolved cases with 10 times finer resolutions to the under-resolved cases with the multi-scale model, comparable global $K_0(\varphi)$, $P_c(Sw)$, and $K_r^\alpha(Sw)$ are obtained [10,12,17]. However, this framework has the following limitations,

I. Since the surface tension effect is already taken into account by the first term in Eq. (1) in the underresolved region, the surface tension originated from the underlying multiphase flow solver must be removed. However, it is not easy to do this with the pseudopotential model without affecting the interface thickness and spurious current. As a result, the surface tension effects are overestimated.

II. After the main interface passes through the under-resolved porous media, some amount of components may be left locally due to the porous media structures. Numerically, however, such residual components cannot be captured with these models because the capillary force acts only in the interface regions. It is also difficult to capture the cell-sized isolated components with the diffusive interface model.

III. The underresolved structures are assumed to be homogeneous structures. Therefore, the permeability in Eq. (1) is the scalar value, although it can be the tensorial form reflecting the directionality of the underresolved porous structure.

In this study, we address such limitations by further refining the numerical models and extending their functionalities to handle various types of multiscale simulations with even higher accuracy. In the following, we first define some problematic cases using the original numerical models. In Section 2, the models that address the problems are presented. In Section 3, they are validated using canonical benchmark cases and practical rock cases. Section 4 summarizes the findings. All quantities are in lattice units unless otherwise stated.

## A. Example case 1

To study the accuracy of the multiscale models, we compare two simulations;
(i) Flow simulation through fully resolved structures with fine resolution
(ii) Flow simulation through the same geometry as in (i), but with much lower resolution, so that some structures are under-resolved. The under-resolved regions are treated with the multiscale models

For a fair comparison, we coarsen the simulation results of (i) with a spatial averaging to compare them with the simulation results of (ii). In all cases in this manuscript, the resolution in the flow simulation is the same as the resolution of the geometry images, where the scanned or modeled images are used to construct the simulation geometry.

An example porous structure of a fully resolved geometry is shown in Figure 2 (a), where the large scale porous media (PM) in the center is sandwiched by a small scale PM structure modified from a model in a resource[14]. The solid material is assumed to be water wet with a contact angle of 10 degrees. The domain size is $1000 \times 200 \times 1000\ pixel$ with an image resolution of $1\mu m/pixel$. We simulate the water imbibition process with the fully resolved image such as Figure 2 (b), which shows the central plane of Figure 2 (a), where black is the fluid and white is the solid. The domain is initially filled with oil. The pressure at the lower boundary is initially 10% higher than that at the upper boundary to prevent the water intrusion, and it is gradually reduced by a small amount so that pressure difference between two boundaries $\Delta P$ is timely varied from 2.5 Pound per Square Inches (PSI) to 0 PSI while a steady state is recorded at each pressure level. The viscosity of both components is $1.66 \times 10^{-3}$. In most of practical cases in this paper, to clarify the correspondence between the physical units and the lattice units, we consider the matching of the Bond number $Bo = \frac{\Delta P L}{\sigma}$ where $L$ is characteristic length and $\sigma$ is the surface tension.

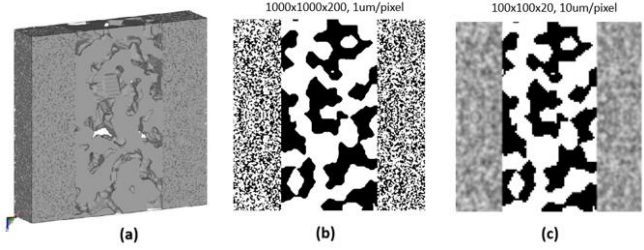

**Figure 2. (a) porous media structures, (b) model image of a central plane at resolution of 1 µm/pixel where black color shows the pore space and white color shows the solid, (c) 10 times coarsened image of a central plane where some of the porous media structure becomes grayscale.**

To simulate with the coarsened model, the input segmented image is first created by coarsening the fully resolved image 10 times, as shown in Figure 2 (c). Then, the small scale PM regions becomes gray, which means containing unresolvable pore sizes. Therefore, it is treated with the multiscale model. The input characteristics are taken from the PM regions, sample of a pore type described in the Appendix section.

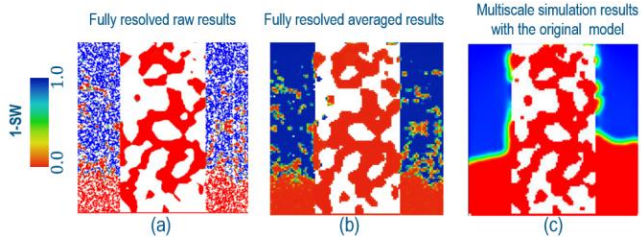

**Figure 3. Water saturation profile during imbibition at $\Delta P =$1.56 PSI. Left is the 1um fully resolved case, center is the 10 times coarsened data by spatially averaging the left result for comparison, right is the multiscale result using models from the previous study[12].**

Figure 3 shows a comparison of the complement of the water saturation, $(1 - Sw)$, during the imbibition process where more than half of the under-resolved region is invaded by water. The three results compared are (a) a simulated result with $1 \mu m/pixel$ fully resolved geometry, (b) 10 times coarsened data by spatially averaging the results in (a) for comparison with multiscale results, and (c) a multiscale simulation result using $10 \mu m/pixel$ coarse geometry with the multiscale numerical models in previous study[12]. The pressure difference between top and bottom boundaries at this moment is 1.56 PSI in the results. Although the critical pressure for invasion appears to be consistent between two cases, it is clear that some of the flow details are not consistent. One inconsistency is the flat invasion front. This is likely due to the overestimated surface tension effects from the underlying multiphase flow solver as discussed in the item I above Second, there is no visible pattern of residual oil trapped in the under-resolved region after the invasion front passes as discussed in the item II above. The detailed irreducible oil patterns are missed here.

## B. Example case 2

In the multiscale model in previous study[12], the under-resolved structure is assumed to be one of the isotropic homogeneous-type porous media. As an example of different anisotropic type of porous media, a bundle of fibers whose flow resistivity has strong directionality is considered. In an in-house made test case, two fiber bundles such as the one shown in Figure 4(b) are placed in a two-dimensional domain as shown in blue in Figure 4(a) where the shape of the fibers is sinusoidal. The color contour shows the porosity distribution, where the red regions represent the fluid cell with $\varphi = 1$, while the blue region is the underresolved PM with $\varphi = 0.95$. Since the fibers are very thin, the porosity variation is small.

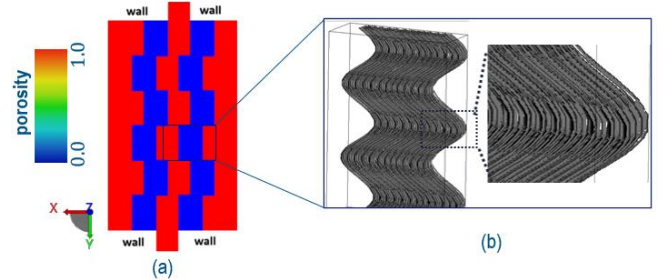

**Figure 4. (a) Porosity distribution of a two-dimensional test case. Red regions have porosity $\varphi = 1$ and blue regions have $\varphi = 0.95$. The top and bottom boundaries are covered with wall and inlet/outlet. (b) an example of a fibrous structure which is assumed to be located in the blue region in (a).**

As a result, the pressure inlet on top boundary and pressure outlet on the bottom boundary are connected by a zigzag pathway. Pressure on the top boundary is 0.740 and on the bottom is 0.733. Left and right boundaries are bounded by walls. Viscosity is $1.66 \times 10^{-2}$. Domain size, excluding inlet/outlet buffer, is $120 \times 180$. The permeability $K_0(\varphi)$ is provided by Figure 19 (a) in Appendix. Although we assume that a bundle of fibers confines the flow to the central path, the simulated result shows much more diffusive flow patterns as shown in Figure 5. This is because the small porosity variation results in a small isotropic resistivity from the under-resolved region. We attempt to remedy this with the tensorial resistivity model and will revisit in Section 3.

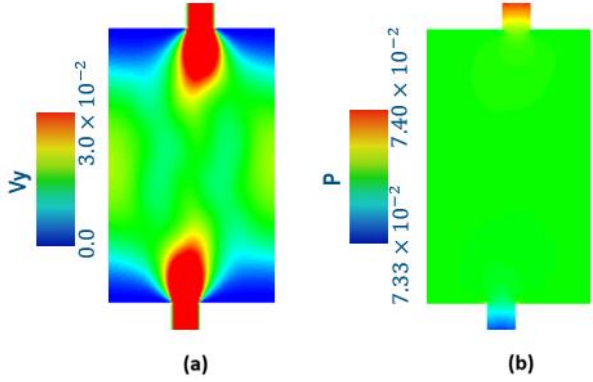

**Figure 5. Multi-scale simulation results using the original numerical model in a system described in Figure 4. (a) Velocity in the y direction and (b) the static pressure.**

## II. METHODOLOGY AND NUMERICAL MODELS

A Lattice Boltzmann (LB) model that has been developed to accurately simulate multicomponent immiscible flow is used in this study. Its inter-component force modeling is based on the Shan-Chan model, the pseudo-potential model[15,16]. The model also incorporates some of the latest advancements for robustness and accuracy[18,19,20,21,22,23,24,25,26,27,28]. In particular, the LB equation can be described as,

$$f_i^\alpha(\vec{x} + \vec{c_i}\Delta t, t + \Delta t) - f_i^\alpha(\vec{x}, t) = \mathcal{C}_i^\alpha + \mathcal{F}_i^\alpha, \quad (2)$$

where $f_i^\alpha$ is the distribution function of fluid component $\alpha$ in the lattice direction $i \in [0,18]$ of the D3Q19 which is the 4th order isotropy lattice model[18], $\vec{c_i}$ is the discrete lattice speed, and $\mathcal{C}_i^\alpha$ is the collision operator projected on the Hermite space[20]. $\mathcal{F}_i^\alpha$ is the force term, which can be written as the following in the case of exact difference method (EDM)[18], for example,

$$\mathcal{F}_i^\alpha = \omega_i \left[ \frac{\overrightarrow{F^\alpha} \cdot \vec{c_i}}{c_s^2} + \frac{(\vec{u}\overrightarrow{F^\alpha} + \overrightarrow{F^\alpha}\vec{u}) : (\vec{c_i}\vec{c_i} - c_s^2 I)}{2c_s^4} \right], \quad (3)$$

where $\omega_i$ is the lattice weights, $\vec{u}$ is the fluid velocity, $c_s$ is speed of sound, and $\overrightarrow{F^\alpha}$ includes various types of forces such as the intercomponent force based on the pseudo potential model[15,16,18], the external force such as the gravity, and the force from the multiscale model such as the one in Eq (1).

To control the surface tension effects on top of the pseudo-potential model, a following CSF (continuous surface force)-type inter-component force[22] is added to $\overrightarrow{F^\alpha}$ of Eq (3),

$$F_{csf,\gamma}^\alpha = \kappa_{csf}(-\nabla \cdot \widehat{n^\alpha}) \frac{\nabla_\gamma \phi^\alpha}{\phi_{avg}^\alpha} \frac{\rho^\alpha}{\phi_{max}^\alpha}, \quad (4)$$

where $\kappa_{csf}$ is a model parameter, $\phi^\alpha = \rho^{\bar{\alpha}}$ is chosen to be the density of the other component than $\alpha$, $\widehat{n^\alpha}$ is the unit vector in the direction of gradient of $\phi^\alpha$, $\phi_{max}^\alpha$ and $\phi_{avg}^\alpha$ are the characteristic maximum and average values of $\phi^\alpha$, respectively. In the multi-scale simulation, non-zero $\kappa_{csf}$ is applied only to the under-resolved PM regions to reduce the surface tension effects.

To capture the residual fluid components in the under-resolved region, we count them separately with the regular fluid components while referring the capillary-pressure-saturation, Pc-Sw, relationship from the input library. Here, for example, a typical simulated Pc-Sw is shown with the red curve in Fig.6. Besides the main capillary pressure features seen around $Sw \approx 0.5$, which are captured with the force model in Eq (1), the residual fluid component observed around $Sw \approx 0.2$ and $Sw \approx 0.9$ is tried to be captured with this new model. Specifically, by defining two reference values, $P_c^{ref,max}$ and $P_c^{ref,min}$, and finding the crossed point with the Pc-Sw curve, shown as the dashed line and yellow and blue dots in Figure 6, the amount of residual components is determined and dynamically stored in the solver. In this study we choose $P_c^{ref,max} = (3.0 \sim 4.0) \cdot P_{c,PM}$ and $P_c^{ref,min} = (0.5 \sim 0.65) \cdot P_{c,PM}$ where $P_{c,PM}$ is the critical pressure in the typical under-resolved region. The yellow dots indicate Sw from residual water and blue dots indicate Sw from residual oil. The Pc-Sw curve for different $\varphi$ can be computed using a simulated data, the red curve, and the scaling law of the Leverett J-function,

$$J = \frac{P_c \sqrt{\frac{K_0}{\varphi}}}{\sigma \cdot \cos\theta}, \quad (5)$$

where $\theta$ is contact angle as shown with the green and blue curves in Figure 6. As shown in the obvious difference among three blue points, the residual component amount can be discontinuously varied from cell to cell.

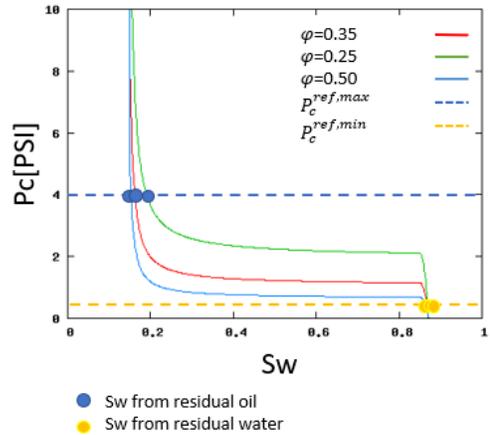

**Figure 6. An example of an input Pc-Sw curve with the red color in the case of $\varphi = 0.35$. The other two lines of green and blue are computed using the red curve and the Leverett J-function similarity for the case of $\varphi = 0.25$ and $\varphi = 0.5$. By finding the crossed points with the input values $P_c^{ref,max}$ and $P_c^{ref,min}$, the amounts of residual components are determined.**

To handle the under-resolved heterogeneous porous media structures, the directionality of the flow resistivity must be accurately accounted for. To detect such a directionality, we refer to the spatial porosity variation and set the flow resistivity tensor along a proper direction as follows. The tensorial resistivity $R_{ij}$ is supposed to be related to the permeability tensor $K_{ij}$ as $R_{ij} = \frac{\nu_{mix}}{K_{ij}}$ where $\nu_{mix}$ is the mixture viscosity[22]. Since $K_{ij}$ is symmetric and positive definite, it can be diagonalized as,

$$K_{ij} = A_{il} K^D_{lm} A^{-1}_{mj}, \quad (6)$$

where $K^D_{lm}$ is the diagonal matrix and $A_{ij}$ is the matrix of eigenvectors of $K_{ij}$. To set $K_{ij}$ along the proper direction, we compute the eigenvectors of Hessian of porosity, $\frac{\partial^2 \varphi}{\partial x_i \partial x_j}$. Then a matrix $B_{il}$ is constructed from them and $A_{il}$ in Eq (6) is replaced by $B_{il}$ as,

$$K_{ij} = B_{il} K^D_{lm} B^{-1}_{mj}. \quad (7)$$

Here the diagonal three elements of $K^D_{lm}$ are inputs. In this way, the permeability tensor $K_{ij}$, and therefore $R_{ij}$, are set along the proper direction from the input diagonal permeability tensor along the principal axis.

## III. VALIDATIONS

### A. A static droplet

A two-dimensional oil droplet, whose radius is 40 lattice, suspended in water, is used to test the surface tension and its effects on the droplet shape, spurious currents, and the interface thickness. In the domain of $100 \times 100$ in size, the periodic boundary condition is imposed on all sides. Figure 7 (a) and (b) show the pressure contour with the model coefficient $\kappa = -0.045$ and $\kappa = 0$. Figure 7 (b) shows that the interior of the droplet has much higher pressure than one in Figure 7 (a), indicating the higher surface tension. Specifically, using the Laplace law, $\sigma = \frac{\Delta P}{R}$, the surface tension $\sigma$ in Figure 7 (b) is calculated to be $0.026$, where $\Delta P$ is the pressure difference between inside and outside the droplet and $R$ is the droplet radius. When the surface tension correction model in Eq (4) is activated, $\sigma$ in Figure 7 (a) becomes $7.43 \times 10^{-4}$, which is reduced by two orders of magnitude, while maintaining the shape of the droplet. Figure 7 (c) shows the centerline plot of the oil density, it shows that the reduced surface tension does not affect the interface thickness very much. Similarly, reduced surface tension does not affect spurious current, either, which is measured by the maximum velocity in steady state and usually occurs along the component interface. The maximum velocity in the x direction is $6.43 \times 10^{-3}$ for $\kappa = 0$, and $6.45 \times 10^{-3}$ for $\kappa = -0.045$.

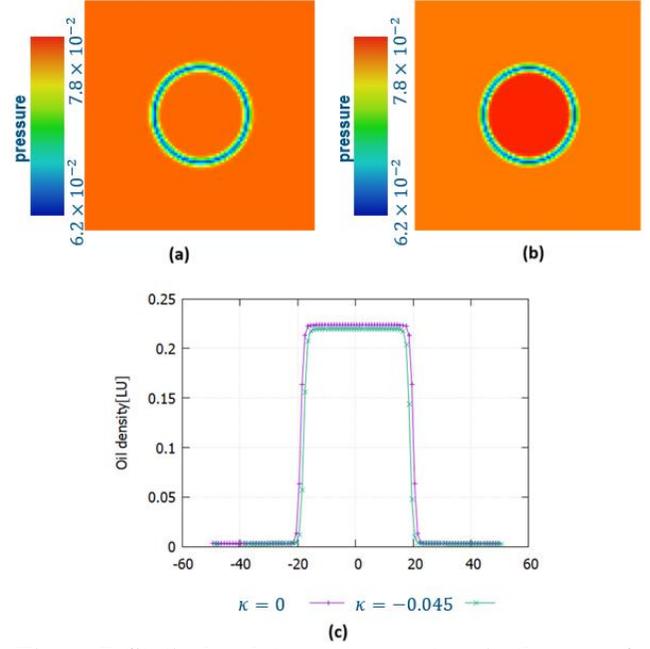

Figure 7. Static droplet pressure contour in the case of (a) $\kappa$=-0.045 and (b) $\kappa$=0. Oil density plot on a center line is shown in (c).

### B. Water imbibition process through random-porosity two-dimensional under-resolved porous media

Water imbibition through under-resolved two-dimensional porous media is simulated. The porosity map is randomly set to follow a normal distribution with a mean of 0.5, as shown in Figure 8.

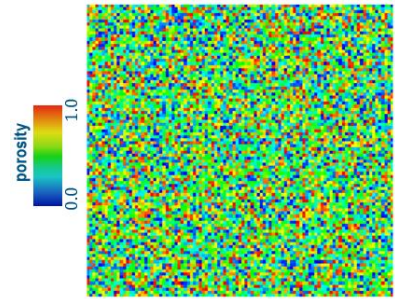

Figure 8. Porosity distribution in a test case in Section III B

The domain size is $100 \times 100$ and is initially filled with oil. The left and right boundaries are solid walls. All walls and under-resolved solids are assumed to be water wet with a contact angle of 10 degrees. The top boundary is the water inlet with 99.0% water saturation and its pressure is set to $7.33 \times 10^{-2}$. The bottom boundary is the oil outlet with 0.05% water saturation. Its pressure is initially set 10% higher than the upper boundary to prevent water intrusion and is gradually reduced as the simulation progresses. Viscosity for both components is set to $1.66 \times 10^{-3}$. The input to the multiscale model uses the constitutive relationships shown in Figure 19 in the Appendix.

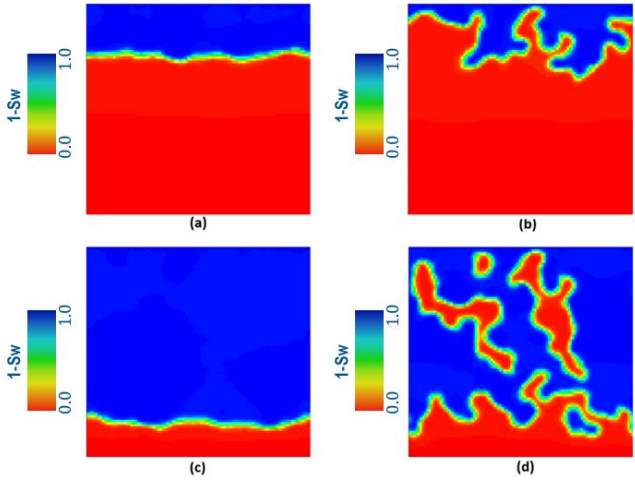

**Figure 9.** Water-oil distribution represented by (1-Sw), i.e. the complement of water saturation at two moments of the imbibition, where blue indicates water and red indicates oil. (a) & (c) Results with the original multiscale model in previous study[12] at early stage and near end of invasion. (b) & (d) Results with the present multiscale model at the early stage and near the end of the invasion.

In Figure 9 (a) and (c), the water-oil distribution at the early and late stages of the invasion is shown with the color contour of $(1 - Sw)$, the complement of water saturation, with the original multiscale model discussed in a previous study[12]. As observed in Section I.A, the invasion front is almost flat except for little unevenness caused by local porosity variation. In Figure 9 (b) and (d) with the reduced surface tension model, the invasion front shows the fingering pattern, which appears to be more realistic. This complex invasion pattern can also form irreducible oil bubbles that remain behind the main interface passing as shown in Figure 9 (d). The capillary number, $Ca = \frac{u\mu}{\sigma}$ where $u$ is the average velocity and $\mu$ is the dynamic viscosity during the invasion is calculated to be $4 \times 10^{-5}$ in the case of Figure 9 (b) and (d). In a previous study[29], at this level of $Ca$ and unity viscosity ratio, the capillary fingering pattern was observed and it seems to be consistent with the results here.

## C. Water imbibition through porous media with large and small scale porous structures

The multiscale models proposed in Sec. II are applied to the case in Sec. IA. The simulation result is compared to one with the original multiscale model and the coarsen one from the fully resolved case with very fine resolution, which does not require the multi-scale model. To illustrate the effects of each modification, we run two cases, one with only the residual component model introduced in Sec. II, and the other with the residual component model plus the reduced surface tension model introduced in Sec. II.

Figure 10 shows the comparison of the imbibition curves. The first large variation, from the left end to Sw ≈ 40% represents the invasion into the small scale porous media (PM) regions. The later large variation around Sw≈48%~80% represents the invasion into the large pore regions. Even with the original multiscale model, these invasion pressures of $P_c \approx 1.5$ PSI and $P_c \approx 0.2$ PSI are quantitatively consistent with the fully resolved case. However, the results with the original model, show the deviation from the fully resolved case with respect to the details of the Sw transition in the intermediate and final stages at $P_c \approx 1.0$ PSI and $P_c \approx 0.0$ PSI. The earlier stage is after the small scale PM regions are fully invaded and before the large scale PM regions are invaded. In the last stage, most of the regions are invaded and the irreducible component contributes to Sw. At these stages, the agreement with the fully resolved case becomes better with the present multiscale model. This is likely because the residual component model accurately captures irreducible components in the small scale PM after the main water interface passes.

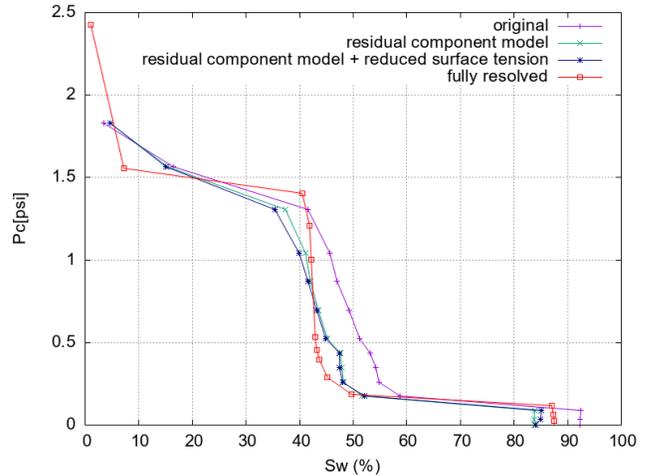

**Figure 10.** Comparison of Pc-Sw imbibition curves among multi-scale simulations using 1) the original multiscale model from previous study[12] and the present multiscale model 2) only with the reduced surface tension model and 3) with the reduced surface tension model and the residual component model. 4) A fully-resolved simulation result with 10times finer resolution without multiscale model is also compared.

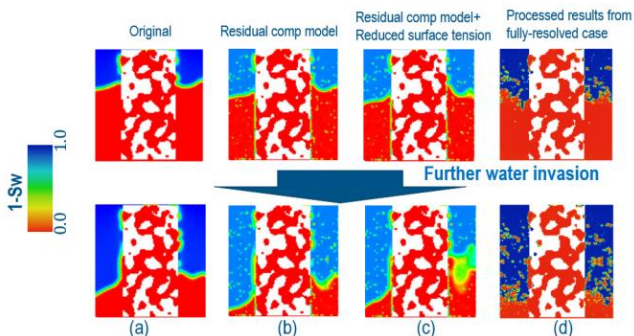

**Figure 11. Comparison of invasion patterns, colored by (1-Sw) using (a) Original multi-scale model in previous study[12], (b) the present multiscale model only with the residual component model, and (c) with the residual component model and the reduced surface tension model. In (d), $10^3$ fluid cells averaged simulation results from a case with 10 times finer resolution and without the multiscale model are compared.**

To see more details of the imbibition patterns, snapshots at two moments are compared in Figure 11. One is near the middle of the small scale PM invasion, the other near its end. Using the original multiscale model in previous study[12], the water invades the under-resolved region with almost flat interface and no oil remains behind the oil/water interface, as shown in Figure 11 (a). When we apply the residual component model in Section II, some of the residual oil can be captured behind the passing main interface, as seen in the dotted light green color in Figure 11 (b). In addition, when we apply the method to reduce the surface tension effects in Sec. II, the main interface shows wavy patterns during the invasion process. Their modifications lead to a better agreement with the simulation results at much higher resolution shown in Figure 11 (d), which does not use the multiscale model. Bond number, at this time instance in the bottom row of Figure 11, is calculated to be 0.096 for all cases.

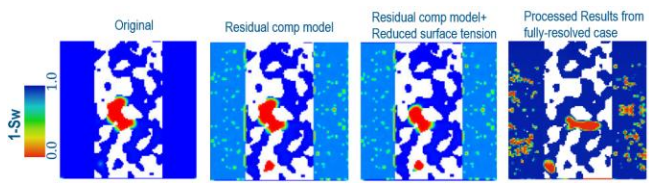

**Figure 12. Comparisons of the irreducible oil distribution at the end of imbibition with the same four options as Figure 11. The trapped large oil blob in the center of the domain is captured in all four cases. The residual component model improves the accuracy of capturing the irreducible components in the small scale PM region.**

At the end of the simulation at $P_c = 0$ PSI, most of the water is swept away throughout the entire domain and only the irreducible components remain, as shown in Figure 12, which compares the same four options as in Figure 11. The large oil blob trapped in the large scale pore space in the center is consistently captured in all cases. As an advantage of the residual component model, much more detailed patterns, with the light green color, are observed in the small scale PM regions as shown in the middle two images of Figure 12. However, compared to the coarsen results of the fully resolved case in the right image of Figure 12, we see some discrepancy in the color range of the irreducible components in the small scale PM region. We consider that one of the reasons is the limitation of the Leverett J-function similarity used in the model, namely its applicability to very different porosity cases.

The reduction in computational cost by using the multi-scale approach is significant when compared with a fully resoved approach. It is a costly task to perform the simulation of the imbibition process for the fully resolved case. The multi-scale approach with the residual component and reduced surface tension model costs 0.3% of computational time of the fully resolved case, which achieves a 3 order-of-magnitude reduction in computational cost. The input characteristics, K0/Pc/Kr profiles, used for the multi-scale model are obtained from simulations in representative sub-domains of the pore type typical porous structures, which accounts for 5% of the computational cost of the fully resolved case. However, this set of profiles only needs to be simulated once, then stored in a library and used in subsequent multi-scale simulations whenever the same type of rock is involved.

### D. Water imbibition to small scale typical carbonate rock

A porous structure as seen in a typical microporous carbonate rock is constructed from open source data[14]. The central plane of three-dimensional model images and its segmented versions, i.e. pixelated image, are shown in Figure 13. The sample size is $256^3$ and the image resolution is $1 \mu m/pixel$. The global porosity is $\varphi_{glb} = 0.38$. Similar to Section III.C, we first run the water imbibition without the multiscale model and then run the multiscale simulation after making the resolution 16 times coarser, namely to $16 \mu m/pixel$, as shown in the right figure of Figure 13. After coarsening, all porous structures are under-resolved and their porosity distribution, which holds $\varphi_{glb} = 0.38$, on a certain plane is shown in Figure 14. Similar to Section III C, the fully resolved simulation results are spatially averaged to allow a fair comparison with the multi-scale case at the resolution of $16 \mu m/pixel$.

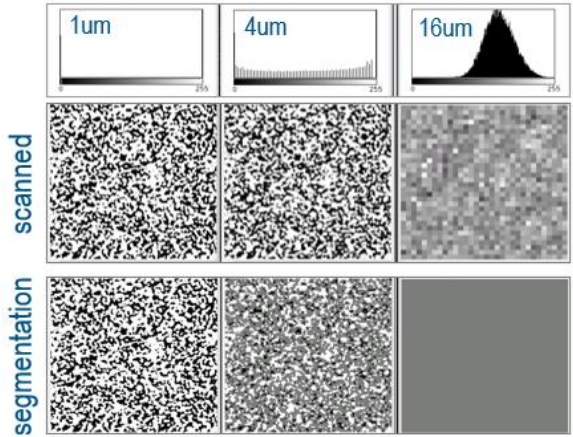

**Figure 13. Coarsening from the original $1\mu m/pixel$ grey scale and segmented images, to $16\mu m/pixel$ (right column) via $4\mu m/pixel$ (middle column). At the last of coarsening, no resolvable pore space is left. Top row is a histogram of pore space distribution with grey scale value of [0:255].**

In the simulation, the viscosity is set as $1.66 \times 10^{-2}$ for both components and all walls including the side-walls and under-resolved solids are assumed to be the water wet of the contact angle 10 degree. The top boundary is water inlet with water saturation of 99.5%, while the bottom boundary is oil outlet with water saturation of 0.5%. For the multiscale simulation, the constitutive relationships in Figure 19 of Appendix are used as inputs. In the fully resolved case, the distribution of the initial fluids is set to locations where they are likely to exist by the pore size analysis[10,17]. On the other hand, in the multi-scale simulation, the oil fills the domain.

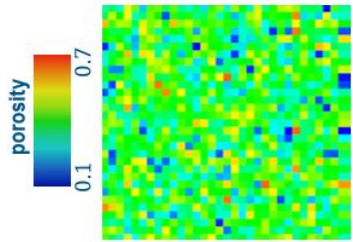

**Figure 14. Porosity distribution in the center plane of the multiscale case. The mean porosity value over the domain is 38.3%.**

In Figure 15, the Pc vs Sw is presented for the multi-scale case with the present model and the fully-resolved case. Since the porosity distribution is smooth around the averaged value as shown in Figure 14, we observe single critical capillary pressure around $P_c \approx 1.5$ PSI. The two simulation results show good agreement.

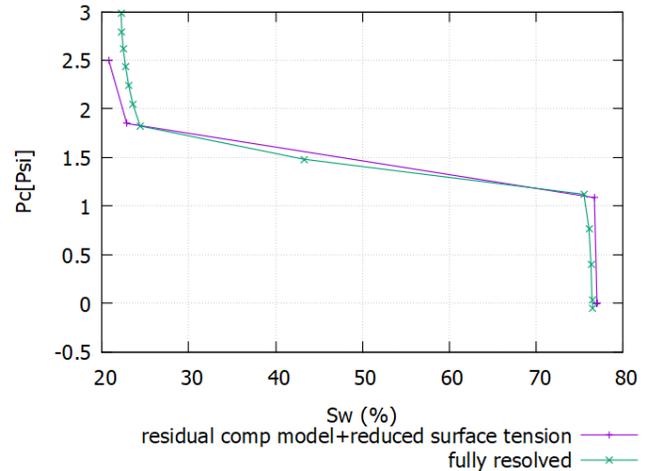

**Figure 15. Comparison of Pc-Sw curve between the fully resolved ($1\mu m/pixel$) case and the multi-scale ($16\mu m/pixel$) case with reduced surface tension model and residual component model.**

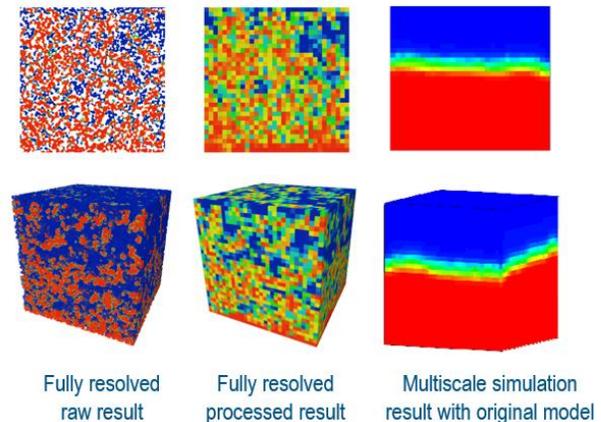

**Figure 16. Simulation results of water saturation where blue is water and red is oil. $1\mu m$ (left) is the direct results from simulation, $16\mu m$ (middle) is 16 times coarsened results from $1\mu m$ and $16\mu m$ (right) is obtained from the previous study[12].**

In Figure 16, the simulation results at $1\mu m/pixel$, their spatially averaged results to $16\mu m/pixel$, and the multiscale simulation results at $16\mu m/pixel$ with the original multiscale model in previous study[12] are compared at two viewing angles. The pressures applied at the moment in Figure 16 are 1.8 PSI for all cases. It indicates that the water invasion occurs with the consistent pressure condition between these cases, and therefore the multiscale model reproduces the reasonable capillary force. However, as shown by the flat color in the right figure in Figure 16, most of the irreducible components are missed. With the present multiscale model, this problem is improved as shown in Figure 17. Here, simulation results with three options are compared at two moments, the early imbibition stage at the top raw and the later imbibition stage at bottom raw. The results from the fully resolved case are shown in (c) and (f). The results with the original multiscale model are shown in (a) and (d). The results with the present multiscale model are shown in (b) and (e). Using the residual component model, we capture the irreducible component distribution with the cell-to-cell variation, which is observed in the fully resolved case. The invasion front interface shape is changed due to lower surface tension. These properties improve the accuracy of multi-scale simulations, making the multi-scale approach a more realistic representation of the fully resolved model. Here Bond number is 0.996 for all cases.

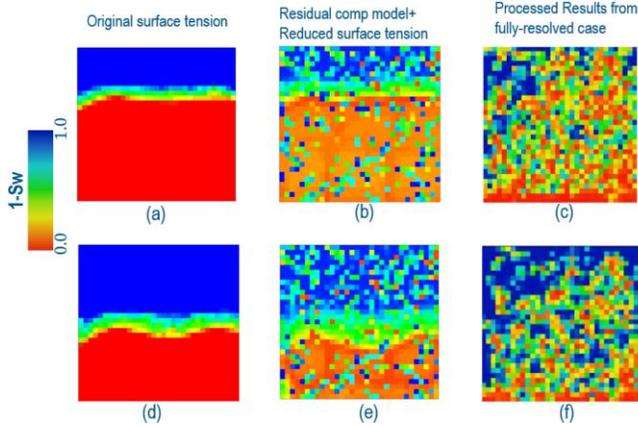

**Figure 17.** Comparisons of invasion patterns, colored by (1-Sw) using (a,d) Original multi-scale model in previous study[12], (b,e) the present multiscale model with the residual component and the reduced surface tension model. In (c,f), averaged simulation results from a case with 16times finer resolution and without the multiscale model are presented. For each option, snapshots at two invasion moments are shown in top and bottom raws.

One of the possible reasons for the different fluid component distributions between the multi-scale case and the fully resolved case is the difference in initialization. In the fully resolved case, the initial component distribution is determined by the pore size analysis, but in the multi-scale simulation, skipping this process, we set the initial fluid distribution mostly from the residual component model. In the future, further optimization of the initial fluid distribution is planned, e.g. by analyzing the connectivity from the porosity distribution.

### E. Single phase flow through a zigzag path separated by a bundle of wavy fibers

The proposed numerical model in Section II is applied to a case introduced in Section IB. Although the scalar resistivity from $K_0(\varphi)$ in Figure 19 (a) is applied in Section IB, the tensorial resistivity is applied so that a least component of the elements of $K_{lm}^D$ in Eq.(7) is set as $10^{-4} K_0(\varphi)$ and the other two components are $K_0(\varphi)$.

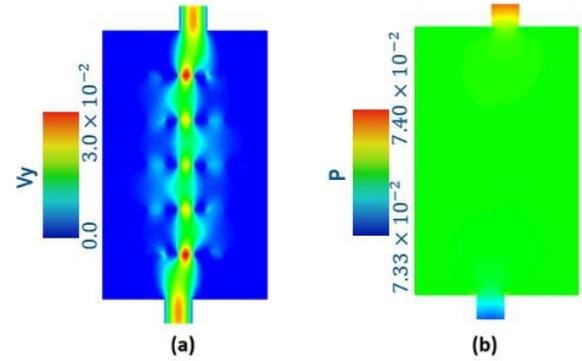

**Figure 18.** (a) velocity in the y direction, (b) static pressure with the multi-scale model of the tensorial resistivity for the comparisons with the results in Figure 5

Figure 18 shows the velocity in the y direction and pressure profiles with the tensorial resistivity. As described in Section II, the tensorial resistivity is constructed from the permeability tensor whose direction is determined by the eigenvectors of the Hessian matrix of the local porosity. As a result, the high resistivity component acts in the right direction and confines the flow to the central passage. The pressure profile also shows that even small changes in porosity shown in Figure 4 result in a barrier to the flow field.

## IV. SUMMARY

Based on the established workflow and multiscale numerical model in a previous study[12], we propose and implement numerical models to improve the accuracy and extend the applications. To define the issues, we set up sample cases and clearly identify the points that need to be addressed. In the test, the compared fine-resolution simulation results show complex behavior, such as bumpy shapes of the component interface as it progresses through the simulation domain, and residual components that are immobile in the small porous media structures. The original multiscale model missed such phenomena because it was designed to capture the fluid force from the under-resolved regions in the leading order. To capture them more accurately, we introduced a correction to the model for reducing the surface tension effects and for capturing irreducible fluid components. In one of test cases of the static droplet, the surface tension is successfully reduced by two orders of magnitude without affecting the spurious current and the interface thickness. In two-dimensional under-resolved porous media, the capillary fingering pattern of the component interface can be seen in agreement with a previous study[29]. In a few test cases based on the rock geometries, the improved models exhibit the complex intrusion patterns and irreducible fluid components in the underresolved regions without significantly changing the main dynamics represented by the Pc-Sw relation. By accurately capturing the irreducible fluid component, the consistency of the detailed Sw value with the fully resolved case improves. To handle different types of under-resolved geometry in the multiscale model, such as one with strong directionality, we formulate a tensorial resistivity whose direction is determined from the local porosity distribution. Through an example case, we show that the model has a significant influence on the flow field and agrees with the expectation. With these promising results, the models and workflows in this study are expected to be applicable to a wide range of engineering systems with multi-scale porous media structures, such as composites, in the future.

## Appendix

In many test cases in this study, the porous structures are created from an open source data[14]. To prepare the input for the multiscale simulation, we performed a pre-simulation for typical domains of this porous structure. Specifically, we randomly picked the domain of $256^3$ with resolution of $1 \mu m/pixel$ from the original data and performed the simulations to compute the absolute permeability $K_0$, capillary pressure $P_c$ and relative permeability $K_r$ as done in the previous study[12]. More details are given below.

Absolute permeability $K_0(\varphi)$ is computed as $\varphi_{glb} < u > v/g$ where $\varphi_{glb} = 0.38$ is the global porosity, $< u >$ is space averaged velocity, and viscosity $v = 0.012$. The acceleration of gravity, $g$, as driving force is $1.0 \times 10^{-4}$. The simulation is set up with a mirrored domain and periodic boundary conditions in the direction of the driving force, while four sides of the domain are covered by solid walls. This type of simulation is performed individually for several samples, all of which are used to interpolate with some standardized analytical forms presented later. The capillary pressure $P_c(S_w)$ is evaluated through imbibition processes, where the domain has an oil reservoir of 14 cell length and the other parts are initially filled by oil. Pressure difference $(P_{top} - P_{bot})$ between top and bottom boundaries is set to sufficiently high initially, then $P_{top}$ is gradually decreased to allow water to invade into the domain. Wettability of all walls are set to be 10 degrees, viscosity of oil/water $v_{water} = v_{oil} = 1.66 \times 10^{-3}$. The simulation for relative permeabilities $K_r^{oil}(Sw)$ and $K_r^{water}(Sw)$ share the same domain setup and driving force as one for $K_0(\varphi)$. Initially, the domain is filled by oil with small amount of water distributed in the small-scale porous media. A controller and mass-sink-source (MSS) function detailed in studies[12,15] are used to determine the steady state and calculate the permeability of both oil/water at each $Sw$ stage and used to inject water around the oil-water interface. Viscosities are set as $v_{water} = v_{oil} = 3.33 \times 10^{-3}$.

The simulated data are fitted with the analytical models such as the Kozney-Carman equation, Thommer model, and the Corey model, for $K_0(\varphi)$, $P_c(Sw)$ and $K_r^{oil/water}(Sw)$, respectively. The functional forms and their coefficient values can be found in Eq. (4.1)-(4.5) in previous study[12]. Results of the fitted curves are shown in Figure 19.

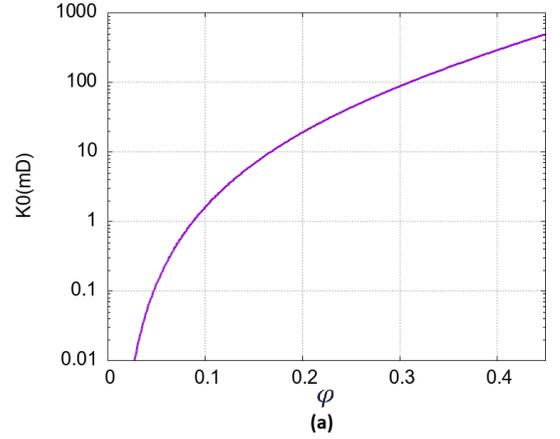

(a)

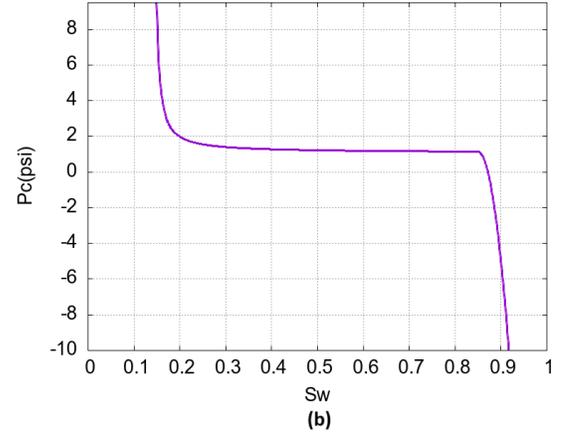

(b)

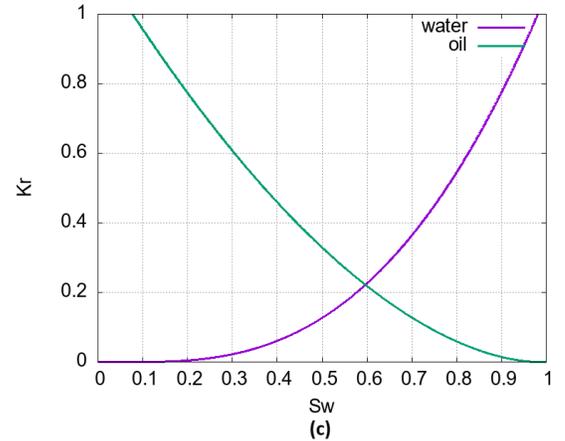

(c)

**Figure 19.** Input data for multi-scale simulations, which are fitted from simulated results (a) absolute permeability vs. porosity, (b) capillary pressure vs. water saturation, (c) effective permeability of oil/water vs. water saturation.